\documentclass[epj]{webofc}

\usepackage[utf8]{inputenc}
\usepackage[varg]{txfonts}   % Web of Conferences font
\usepackage{booktabs}
\usepackage{xcolor}
\definecolor{darkred}{rgb}{0.4,0.0,0.0}
\definecolor{darkgreen}{rgb}{0.0,0.4,0.0}
\definecolor{darkblue}{rgb}{0.0,0.0,0.4}

\usepackage[bookmarks,linktocpage,colorlinks,
    linkcolor = darkred,
    urlcolor  = darkblue,
    citecolor = darkgreen]{hyperref}

\usepackage{fancyvrb} %%% needed to use colors in "verbatim" environment
\definecolor{myred}{rgb}{0.9,0.0,0.0}
\definecolor{mygreen}{rgb}{0.2,0.6,0.2}
\definecolor{myblue}{rgb}{0.0,0.0,0.9}
\wocname{EPJ Web of Conferences}
\woctitle{Lattice2017}

\newcommand{\ga}{\gamma}
\newcommand{\de}{\delta}

\newcommand{\la}{\lambda}
\newcommand{\rh}{\rho}

\newcommand{\si}{\sigma}

\renewcommand{\dag}{^\dagger}

\newcommand{\nab}{\nabla}
\newcommand{\lap}{\triangle}

\newcommand{\bdm}{\begin{displaymath}}
\newcommand{\edm}{\end{displaymath}}
\newcommand{\bea}{\begin{eqnarray}}
\newcommand{\eea}{\end{eqnarray}}
\newcommand{\beq}{\begin{equation}}
\newcommand{\eeq}{\end{equation}}

\newcommand{\mr}{\mathrm}

\newcommand{\ri}{\mathrm{i}}

\newcommand{\Nc}{N_{\!c}}
\newcommand{\Nv}{N_{\!v}}

\begin{document}

\selectlanguage{english}
%----------------------------------------------------------------------------
\title{%
Optimization of the Brillouin operator on the KNL architecture
}
%----------------------------------------------------------------------------
\author{%
\firstname{Stephan} \lastname{D\"urr}\inst{1,2}
}
%----------------------------------------------------------------------------
\institute{%
University of Wuppertal, Gau{\ss}stra{\ss}e 20, D-42119 Wuppertal, Germany
\and
IAS/JSC, Forschungszentrum J\"ulich GmbH, D-52425 J\"ulich, Germany
}
%----------------------------------------------------------------------------
\abstract{%
Experiences with optimizing the matrix-times-vector application of the Brillouin operator on the Intel KNL processor are reported.
Without adjustments to the memory layout, performance figures of 360 Gflop/s in single and 270 Gflop/s in double precision are observed.
This is with $N_\mr{c}\!=\!3$ colors, $N_\mr{v}\!=\!12$ right-hand-sides, $N_\mr{thr}\!=\!256$ threads, on lattices of size $32^3\times64$, using exclusively OMP pragmas.
Interestingly, the same routine performs quite well on Intel Core i7 architectures, too.
Some observations on the much harder Wilson fermion matrix-times-vector optimization problem are added.
}
%----------------------------------------------------------------------------
\maketitle

%%%%%%%%%%%%%%%%%%%%%%%%%%%%%%%%%%%%%%%%%%%%%%%%%%%%%%%%%%%%%%%%%%%%%%%%%%%%%

\section{Introduction}\label{introduction}

%\begin{table}[tb]
%  \small
%  \centering
%  \caption{Please write your table caption here}
%  \label{tab-1}% Give a unique label
%  \begin{tabular}{lll}       \toprule
%  first  & second & third  \\\midrule
%  number & number & number \\
%  number & number & number \\\bottomrule
%  \end{tabular}
%\end{table}

Conceptually the Brillouin operator $D_\mr{B}$ is a sibling of the Wilson Dirac operator $D_\mr{W}$, since
\bea
D_\mr{W}(x,y)&=&\sum_\mu \ga_\mu \nab_\mu^\mr{std}(x,y)
-\frac{a}{2}\lap^\mr{std}(x,y)+m_0\de_{x,y}
-\frac{c_{{}_\mr{SW}}}{2}\sum_{\mu<\nu}\si_{\mu\nu}F_{\mu\nu}\de_{x,y}
\label{def_W}
\\
D_\mr{B}(x,y)&=&\sum_\mu \ga_\mu \nab_\mu^\mr{iso}(x,y)
-\frac{a}{2}\lap^\mr{bri}(x,y)+m_0\de_{x,y}
-\frac{c_{{}_\mr{SW}}}{2}\sum_{\mu<\nu}\si_{\mu\nu}F_{\mu\nu}\de_{x,y}
\label{def_B}
\eea
share the same structure%
\footnote{$m_0$ and $c_{{}_\mr{SW}}$ must be tuned separately in (\ref{def_W}) and (\ref{def_B}) to establish vanishing pion mass and absence of $O(a)$ cut-off effects.}
with $\si_{\mu\nu}\!=\!\frac{\ri}{2}[\ga_\mu,\ga_\nu]$ and $F_{\mu\nu}$ the hermitean clover-leaf field-strength tensor.
The main difference is that the isotropic derivative $\nab_\mu^\mr{iso}$ and the Brillouin Laplacian $\lap^\mr{bri}$ both include 80 neighbors (all within the $[-1,+1]^4$ hypercube
around a given lattice point $x$, i.e.\ up to 4 hops away from $x$, but no two hops may be in the same direction), rather than the 8 neighbors present in the standard derivative $\nab_\mu^\mr{std}$ and the standard Laplacian $\lap^\mr{std}$.

Given that the Brillouin operator has a larger ``footprint'' and hence more operations per site
than the Wilson operator, a natural question to ask is whether $D_\mr{B}$ is more suitable for modern architectures (which typically involve lots of cores, but are limited by memory bandwidth) than  $D_\mr{W}$.

As a first step to address this question, I decided to come up with a \emph{simple} implementation of the Brillouin operator on the Intel KNL architecture.
More specifically, this boundary condition is meant to imply that ($i$) only shared memory parallelization via OpenMP pragmas on a single CPU is used (i.e.\ no distributed memory parallelization with MPI), ($ii$) the memory layout is unchanged from the generic layout used throughout my code suite (see below for details), and ($iii$) no single-thread performance tuning is attempted beyond adding straightforward SIMD hints (again via OpenMP pragmas).
To the expert these constraints may look unnecessarily tight, but it turns out that nonetheless sustained performance figures%
\footnote{In the meantime (i.e.\ after the conference) a slight increase to 380 Gflop/s was achieved.}
of 360 Gflop/s in single precision (sp) arithmetics can be achieved.

%%%%%%%%%%%%%%%%%%%%%%%%%%%%%%%%%%%%%%%%%%%%%%%%%%%%%%%%%%%%%%%%%%%%%%%%%%%%%

\section{Brillouin operator in a nutshell}\label{nutshell}

Let $(\la_0,\la_1,\la_2,\la_3,\la_4)\equiv(-240,8,4,2,1)/64$; then the free $\lap^\mr{bri}$ in (\ref{def_B}) takes the form
\bea
 a^2\lap^\mr{bri}(x,y)&=&\la_0\,\de_{x,y}
                       + \la_1\sum\nolimits_{\mu}\de_{x+\hat\mu,y}
                       + \la_2\sum\nolimits_{\neq(\nu,\mu)}\de_{x+\hat\mu+\hat\nu,y}
\nonumber\\
                      &+&\la_3\sum\nolimits_{\neq(\rh,\nu,\mu)}\de_{x+\hat\mu+\hat\nu+\hat\rh,y}
                       + \la_4\sum\nolimits_{\neq(\si,\rh,\nu,\mu)}\de_{x+\hat\mu+\hat\nu+\hat\rh+\hat\si,y}
\label{def_laplace}
\eea
where $\neq\!(\rh,\nu,\mu)$ means that $\rh$, $\nu$ and $\mu$ are summed over, subject to the constraint that no two elements are equal.
Similarly, with $(\rh_1,\rh_2,\rh_3,\rh_4)\equiv(64,16,4,1)/432$ the free $\nab_\mu^\mr{iso}$ in (\ref{def_B}) takes the form
\bea
 a\nab_\mu^\mr{iso}(x,y)&=&\rh_1\,[\de_{x+\hat\mu,y}-\de_{x-\hat\mu,y}]
\nonumber\\
&+&\rh_2\sum\nolimits_{\neq(\nu;\mu)}[\de_{x+\hat\mu+\hat\nu,y}-\de_{x-\hat\mu+\hat\nu,y}]
\nonumber\\
&+&\rh_3\sum\nolimits_{\neq(\rh,\nu;\mu)}[\de_{x+\hat\mu+\hat\nu+\hat\rh,y}-\de_{x-\hat\mu+\hat\nu+\hat\rh,y}]
\nonumber\\
&+&\rh_4\sum\nolimits_{\neq(\si,\rh,\nu;\mu)}[\de_{x+\hat\mu+\hat\nu+\hat\rh+\hat\si,y}-\de_{x-\hat\mu+\hat\nu+\hat\rh+\hat\si,y}]
\label{def_nabla}
\eea
where $\neq\!(\rh,\nu;\mu)$ means that only $\rh$ and $\nu$ are summed over, while still no two out of the three indices may be equal.
In the interacting theory, these stencils are to be gauged in the obvious way.
For instance, considering (\ref{def_laplace}) we see 8 terms ($\propto\!\la_1$) with 1 hop; they are dressed with $U_\mu(x)$ or $U_\mu(x\!-\hat\mu)\dag$, for positive or negative $\mu$, respectively.
Similarly, there are 24 terms with 2 hops; they are dressed with off-axis links of the form $\frac{1}{2}[U_\mu(x)U_\nu(x\!+\!\hat\mu)+U_\nu(x)U_\mu(x\!+\!\hat\nu)]$.
Next, there are 32 terms with 3 hops; they are dressed with off-axis links which are the average of 6 products of three factors each.
And finally, there are 16 terms with 4 hops; they are dressed with hyperdiagonal links built as the average of 24 products of four factors each.
Further details of the operator can be found in \cite{Durr:2010ch,Durr:2017wfi}.

%%%%%%%%%%%%%%%%%%%%%%%%%%%%%%%%%%%%%%%%%%%%%%%%%%%%%%%%%%%%%%%%%%%%%%%%%%%%%

\section{Code suite overall guidelines}\label{guidelines}

The overall guidelines of the code suite are best illustrated by taking a look at the Wilson operator
\beq
D_\mr{W}(x,y)=\frac{1}{2}\sum_\mu
\Big\{
(\ga_\mu\!-\!I) \; U_\mu(x)                 \; \de_{x+\hat\mu,y}-
(\ga_\mu\!+\!I) \; U_\mu\dag(x\!-\!\hat\mu) \; \de_{x-\hat\mu,y}
\Big\}
+(4\!+\!m_0) \; \de_{x,y}
\label{wilson}
\eeq
which is said to operate on a vector of length $\Nc 4 N_x N_y N_z N_t$, where $\Nc$ is the number of colors.

From a computational viewpoint it is extremely convenient to declare the source and sink vectors as complex arrays of size \texttt{(1:Nc,1:4,1:Nv,1:Nx*Ny*Nz*Nt)}.
Here the ordering and the stride notation common to Matlab and Fortran are used; with the default counting from 1 we could write \texttt{(Nc,4,Nv,Nx*Ny*Nz*Nt)}.
A special feature is that we have one slot to address the right-hand-side (rhs), i.e.\ $\Nv$ columns (in the mathematical setup terminology) can be processed simultaneously.
The underlying philosophy is that index computations are done by the compiler, except for the site index \texttt{n=(l-1)*Nx*Ny*Nz+(k-1)*Nx*Ny+(j-1)*Nx+i}, where we want
to keep some freedom.

\begin{figure}[tb]
\begin{Verbatim}[fontsize=\footnotesize,commandchars=\\\{\}]
PARAMETERS: Nx,Ny,Nz,Nt, Nc,Nv, sp,dp, i_sp,i_dp {\color{mygreen}!!! known at compile time}
{\color{myred}!$OMP PARALLEL DO DEFAULT(private) FIRSTPRIVATE(mass) SHARED(old,new,V) SCHEDULE(static)}
{\color{myblue}do} l=1,Nt; l_plu=modulo(l,Nt)+1; l_min=modulo(l-2,Nt)+1
{\color{myblue}do} k=1,Nz; k_plu=modulo(k,Nz)+1; k_min=modulo(k-2,Nz)+1
{\color{myblue}do} j=1,Ny; j_plu=modulo(j,Ny)+1; j_min=modulo(j-2,Ny)+1
{\color{myblue}do} i=1,Nx; i_plu=modulo(i,Nx)+1; i_min=modulo(i-2,Nx)+1
   n=(((l-1)*Nz+(k-1))*Ny+(j-1))*Nx+i {\color{mygreen}!!! volume index}
   {\color{mygreen}!!! direction 0 gets factor -1/2*(-8I)=4*I and mass}
   site(:,:,:)=(4.0+mass)*old(:,:,:,n)
   {\color{mygreen}!!! direction -4 gets factor 1/2*(-I-gamma4)}
   tmp=0.5*transpose(conjg(V(:,:,4,i,j,k,l_min))); nsh=n+(l_min-l)*Nz*Ny*Nx; ...
   {\color{mygreen}!!! direction -3 gets factor 1/2*(-I-gamma3)}
   tmp=0.5*transpose(conjg(V(:,:,3,i,j,k_min,l))); nsh=n+(k_min-k)*Ny*Nx   ; ...
   {\color{mygreen}!!! direction -2 gets factor 1/2*(-I-gamma2)}
   tmp=0.5*transpose(conjg(V(:,:,2,i,j_min,k,l))); nsh=n+(j_min-j)*Nx      ; ...
   {\color{mygreen}!!! direction -1 gets factor 1/2*(-I-gamma1)}
   tmp=0.5*transpose(conjg(V(:,:,1,i_min,j,k,l))); nsh=n+(i_min-i)         ; ...
   {\color{mygreen}!!! direction +1 gets factor 1/2*(-I+gamma1)}
   tmp=0.5*                V(:,:,1,i    ,j,k,l)  ; nsh=n+(i_plu-i)         ; ...
   {\color{mygreen}!!! direction +2 gets factor 1/2*(-I+gamma2)}
   tmp=0.5*                V(:,:,2,i,j    ,k,l)  ; nsh=n+(j_plu-j)*Nx      ; ...
   {\color{mygreen}!!! direction +3 gets factor 1/2*(-I+gamma3)}
   tmp=0.5*                V(:,:,3,i,j,k    ,l)  ; nsh=n+(k_plu-k)*Ny*Nx   ; ...
   {\color{mygreen}!!! direction +4 gets factor 1/2*(-I+gamma4)}
   tmp=0.5*                V(:,:,4,i,j,k,l    )  ; nsh=n+(l_plu-l)*Nz*Ny*Nx; ...
{\color{myblue}end do} ! i=1,Nx
{\color{myblue}end do} ! j=1,Ny
{\color{myblue}end do} ! k=1,Nz
{\color{myblue}end do} ! l=1,Nt
{\color{myred}!$OMP END PARALLEL DO}
\end{Verbatim}
\vspace*{-9pt}%
\caption{Overall structure of the Wilson routine; accumulation of the 1+8 contributions in the thread-private
variable \texttt{site(1:Nc,1:4,1:Nv)} in the dotted blocks is specified in Fig.\,\ref{fig:wils_detail}.
In the actual routine \texttt{COLLAPSE(2)} is added to the \texttt{!\$OMP} pragma, and the the statements
for \texttt{l\_plu} and \texttt{l\_min} are transferred to the next line.}%
\label{fig:wils_overall}%
\end{figure}

\begin{figure}[tb]
\begin{Verbatim}[fontsize=\footnotesize,commandchars=\\\{\}]
{\color{mygreen}!!! direction -1 gets factor 1/2*(-I-gamma1)}
tmp=0.5*transpose(conjg(V(:,:,1,i_min,j,k,l))); nsh=n+(i_min-i)
{\color{myred}!$OMP SIMD PRIVATE(full)}
{\color{myblue}do} idx=1,Nv
   {\color{myblue}forall}(col=1:Nc,spi=1:4) full(col,spi)=sum(tmp(col,:)*old(:,spi,idx,nsh))
   site(:,1,idx)=site(:,1,idx)-full(:,1)+i_sp*full(:,4) {\color{mygreen}!!! gamma1^trsp=    0     0     0     i}
   site(:,2,idx)=site(:,2,idx)-full(:,2)+i_sp*full(:,3) {\color{mygreen}!!!                 0     0     i     0}
   site(:,3,idx)=site(:,3,idx)-full(:,3)-i_sp*full(:,2) {\color{mygreen}!!!                 0    -i     0     0}
   site(:,4,idx)=site(:,4,idx)-full(:,4)-i_sp*full(:,1) {\color{mygreen}!!!                -i     0     0     0}
{\color{myblue}end do}
{\color{myred}!$OMP END SIMD}
\end{Verbatim}
\vspace*{-9pt}%
\caption{Detail of the fourth block in Fig.\,\ref{fig:wils_overall}; the variable \texttt{full(1:Nc,1:4)} contains \texttt{old(1:Nc,1:4,idx,nsh)}
after left-multiplication with $\frac{1}{2}$ times the link-variable, but before right-multiplication with $(-I-\gamma_1)^\mathrm{trsp}$.}%
\label{fig:wils_detail}%
\end{figure}

The working of these guidelines, on the basis of the Wilson operator (\ref{wilson}), is spelled out in the code assembled in Figs.\,\ref{fig:wils_overall} and \ref{fig:wils_detail}.
The arrangement of the loops over the $x,y,z,t$ coordinates (denoted \texttt{i,j,k,l}, respectively) makes sure the innermost loop belongs to the fastest index.
A noteworthy mathematical detail is how the $\Nc4\times\Nc4$ matrix $\frac{1}{2}(\ga_\mu\!-\!I) \otimes U_\mu(x)$ acts on the $\Nc4\times1$ vector \texttt{old(:,:,idx,n)} with given rhs and site indices.
This product is realized by reshaping it into a $\Nc\times4$ matrix (which it already is in our setup), multiplying it with $U_\mu(x)$ from the left, and with $\frac{1}{2}(\ga_\mu\!-\!I)^\mr{trsp}$ from the right.
The full-spinor variable \texttt{full(Nc,4)} contains the result of the former multiplication; the $\gamma$-operation just reorders its columns (modulo some signs and factors of $\mr{i}$).

The name of the gauge variable in Figs.\,\ref{fig:wils_overall} and \ref{fig:wils_detail} is supposed to remind us that in most practical applications the smeared gauge field $V_\mu(x)$ is used
rather than the original gauge field $U_\mu(x)$.
In case of clover improvement, it is practical to precompute the field-strength tensor $F_{\mu\nu}(x)$ once, and to store it in a complex array \texttt{F(Nc,Nc,6,Nx,Ny,Nz,Nt)} which is then passed to the clover routine.
One might notice that in Fig.\,\ref{fig:wils_detail} only the linear combinations \texttt{full(:,1)-i*full(:,4)} and \texttt{full(:,2)-i*full(:,3)} are used.
Hence it would have been sufficient to form these linear combinations prior to left-multiplying with the gauge field.
This ``shrink-expand-trick'' can be used for all eight directions, regardless of the $\gamma$-representation chosen (Fig.\,\ref{fig:wils_detail} uses the chiral one).

%%%%%%%%%%%%%%%%%%%%%%%%%%%%%%%%%%%%%%%%%%%%%%%%%%%%%%%%%%%%%%%%%%%%%%%%%%%%%

\section{Brillouin kernel details}\label{details}

\begin{figure}[tb]
\begin{Verbatim}[fontsize=\footnotesize,commandchars=\\\{\}]
PARAMETERS: Nx,Ny,Nz,Nt, Nc,Nv, sp,dp, i_sp,i_dp {\color{mygreen}!!! known at compile time}
parameter(kind=sp),dimension(0:4) :: mask=[0.0_sp,64.0_sp,16.0_sp,4.0_sp,1.0_sp]/432.0_sp
{\color{myred}!$OMP PARALLEL DO COLLAPSE(2) FIRSTPRIVATE(mass) SHARED(old,new,W) SCHEDULE(static)}
{\color{myblue}do} l=1,Nt
{\color{myblue}do} k=1,Nz
{\color{myblue}do} j=1,Ny
{\color{myblue}do} i=1,Nx
   n=(((l-1)*Nz+(k-1))*Ny+(j-1))*Nx+i {\color{mygreen}!!! volume index}
   site(:,:,:)=cmplx(0.0,kind=sp) {\color{mygreen}!!! note: site is Nc*4*Nv}
   {\color{mygreen}!!! visit all 81 sites within hypercube (distances 0 to 4 in taxi-driver metric)}
   dir=0
   {\color{myblue}do} go_l=-1,1; lsh=modulo(l+go_l-1,Nt)+1
   {\color{myblue}do} go_k=-1,1; ksh=modulo(k+go_k-1,Nz)+1
   {\color{myblue}do} go_j=-1,1; jsh=modulo(j+go_j-1,Ny)+1
   {\color{myblue}do} go_i=-1,1; ish=modulo(i+go_i-1,Nx)+1
      dir=dir+1 {\color{mygreen}!!! note: dir=(go_l+1)*27+(go_k+1)*9+(go_j+1)*3+go_i+2}
      ...
   {\color{myblue}end do} ! go_i=-1,1
   {\color{myblue}end do} ! go_j=-1,1
   {\color{myblue}end do} ! go_k=-1,1
   {\color{myblue}end do} ! go_l=-1,1
   {\color{mygreen}!!! plug everything into new vector}
   {\color{myblue}do} idx=1,Nv
      new(:,:,idx,n)=site(:,:,idx)
   {\color{myblue}end do} ! idx=1,Nv
{\color{myblue}end do} ! i=1,Nx
{\color{myblue}end do} ! j=1,Ny
{\color{myblue}end do} ! k=1,Nz
{\color{myblue}end do} ! l=1,Nt
{\color{myred}!$OMP END PARALLEL DO}
\end{Verbatim}
\vspace*{-9pt}%
\caption{Overall structure of the Brillouin routine; accumulation of the 81 contributions to the thread-private variable \texttt{full(1:Nc,1:4)} happens in
four inner loops with details specified in Fig.\,\ref{fig:bril_detail}.}%
\label{fig:bril_overall}%
\end{figure}

\begin{figure}[tb]
\footnotesize
\begin{Verbatim}[fontsize=\footnotesize,commandchars=\\\{\}]
select case(dir)
   case(01:40); tmp=W(:,:,dir,i,j,k,l)
   case(   41); tmp=color_eye() {\color{mygreen}!!! note: yields Nc*Nc identity matrix}
   case(42:81); tmp=conjg(transpose(W(:,:,82-dir,ish,jsh,ksh,lsh)))
end select
absgo_ijkl=sum([go_i,go_j,go_k,go_l]**2) {\color{mygreen}!!! go_i**2+go_j**2+go_k**2+go_l**2}
fac=0.125_sp/2**absgo_ijkl {\color{mygreen}!!! note: factor for 1/2 times Brillouin Laplacian}
if (absgo_ijkl.eq.0) fac=fac-2.0_sp-mass {\color{mygreen}!!! note: correction for go_i=go_j=go_k=go_l=0}
fac_i=go_i*mask(absgo_ijkl) {\color{mygreen}!!! note: factor for isotropic derivative in x-direction}
fac_j=go_j*mask(absgo_ijkl) {\color{mygreen}!!! note: factor for isotropic derivative in y-direction}
fac_k=go_k*mask(absgo_ijkl) {\color{mygreen}!!! note: factor for isotropic derivative in z-direction}
fac_l=go_l*mask(absgo_ijkl) {\color{mygreen}!!! note: factor for isotropic derivative in t-direction}
nsh=(((lsh-1)*Nz+(ksh-1))*Ny+(jsh-1))*Nx+ish
{\color{myred}!$OMP SIMD PRIVATE(full)}
{\color{myblue}do} idx=1,Nv
   {\color{myblue}forall}(col=1:Nc,spi=1:4) full(col,spi)=sum(tmp(col,:)*old(:,spi,idx,nsh))
   site(:,:,idx)=site(:,:,idx)-fac*full(:,:)
   site(:,1,idx)=site(:,1,idx)+cmplx(-fac_j,-fac_i)*full(:,4)+cmplx(+fac_l,-fac_k)*full(:,3)
   site(:,2,idx)=site(:,2,idx)+cmplx(+fac_j,-fac_i)*full(:,3)+cmplx(+fac_l,+fac_k)*full(:,4)
   site(:,3,idx)=site(:,3,idx)+cmplx(+fac_j,+fac_i)*full(:,2)+cmplx(+fac_l,+fac_k)*full(:,1)
   site(:,4,idx)=site(:,4,idx)+cmplx(-fac_j,+fac_i)*full(:,1)+cmplx(+fac_l,-fac_k)*full(:,2)
{\color{myblue}end do} ! idx=1,Nv
{\color{myred}!$OMP END SIMD}
\end{Verbatim}
\vspace*{-9pt}%
\caption{Detail of the dotted block in Fig.\,\ref{fig:bril_overall}; the variable \texttt{full(1:Nc,1:4)} contains \texttt{old(1:Nc,1:4,idx,nsh)}
after left-multiplication with the (off-axis) link-variable, but before right-multiplication with the $\gamma$-structure.}%
\label{fig:bril_detail}%
\end{figure}

The Brillouin matrix-times-vector routine, built according to the guidelines laid out in the previous section, is portrayed in Figs.\,\ref{fig:bril_overall} and \ref{fig:bril_detail}.
The four-fold loop structure over the out-vector is OMP-parallelized in the a straightforward way (the \texttt{COLLAPSE(2)} statement makes sure that up to $N_zN_t$ threads can be launched).
The most important difference to the Wilson routine is that 40 out of the 81 directions of the off-axis links $W$, build from the smeared gauge field $V$, are assembled in the
complex array \texttt{W(Nc,Nc,40,Nx,Ny,Nz,Nt)}; the remaining ones are the identity or the hermitean conjugate of $W$ in a hypercube related point (i.e.\ up to four hops away).

Similar to the Wilson routine, the SIMD hints are given as pragmas to the loop over the rhs-index \texttt{idx}.
Within this loop, the spinor and color operations are explicitly or implicitly unrolled (e.g.\ by forall constructs, stride notation).
The complex numbers formed from \texttt{fac\_i,fac\_j,fac\_k,fac\_l} implement the right-multiplication with $\gamma_1^\mr{trsp},...,\gamma_4^\mr{trsp}$ in the chiral representation.
All 81 contributions are accumulated in the thread-private variable \texttt{site(Nc,4,Nv)};
since this variable is written once to the respective site in \texttt{new(:,:,:,n)} there cannot be any thread-write-collision by construction.

%%%%%%%%%%%%%%%%%%%%%%%%%%%%%%%%%%%%%%%%%%%%%%%%%%%%%%%%%%%%%%%%%%%%%%%%%%%%%

\section{Brillouin operator timings}\label{bril-timings}

Timings are done on a node containing a single KNL chip with 64 cores.
All results for the Brillouin operator are converted into Gflop/s, based on a flop count of $2560\Nc^2+2376\Nc$ per site (i.e.\ 30168 for QCD, see \cite{Durr:2017wfi} for details).
As a default setup we shall use a $32^3\times64$ lattice, with $\Nc=3$ and $\Nv=4\Nc$.
This geometry is chosen such that a sp-field \texttt{W}, a sp-field \texttt{F}, and the dp-fields \texttt{old,new} fit into the high-bandwidth MCDRAM of 16\,GB.
For any number of threads, memory allocation of these objects to the individual cores is done by a first-touch policy.
The static thread scheduling makes sure every thread gets exactly the same fraction of the \texttt{out} vector to work on.
Compilation is done with ifort version 17.2, with the flags \texttt{-qopenmp -O2 -xmic-avx512 -align array64byte}.

\begin{figure}[tb]
\includegraphics[width=0.5\textwidth]{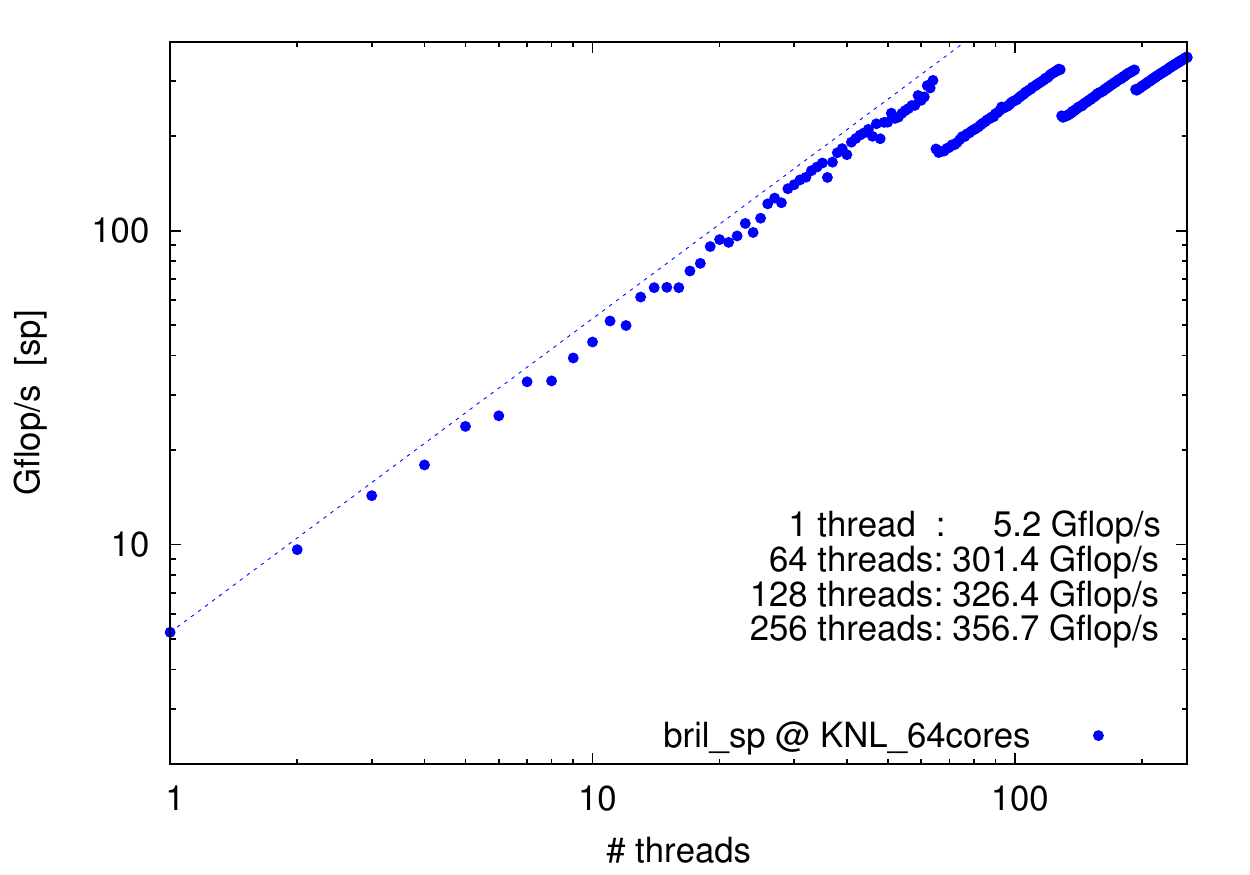}%
\includegraphics[width=0.5\textwidth]{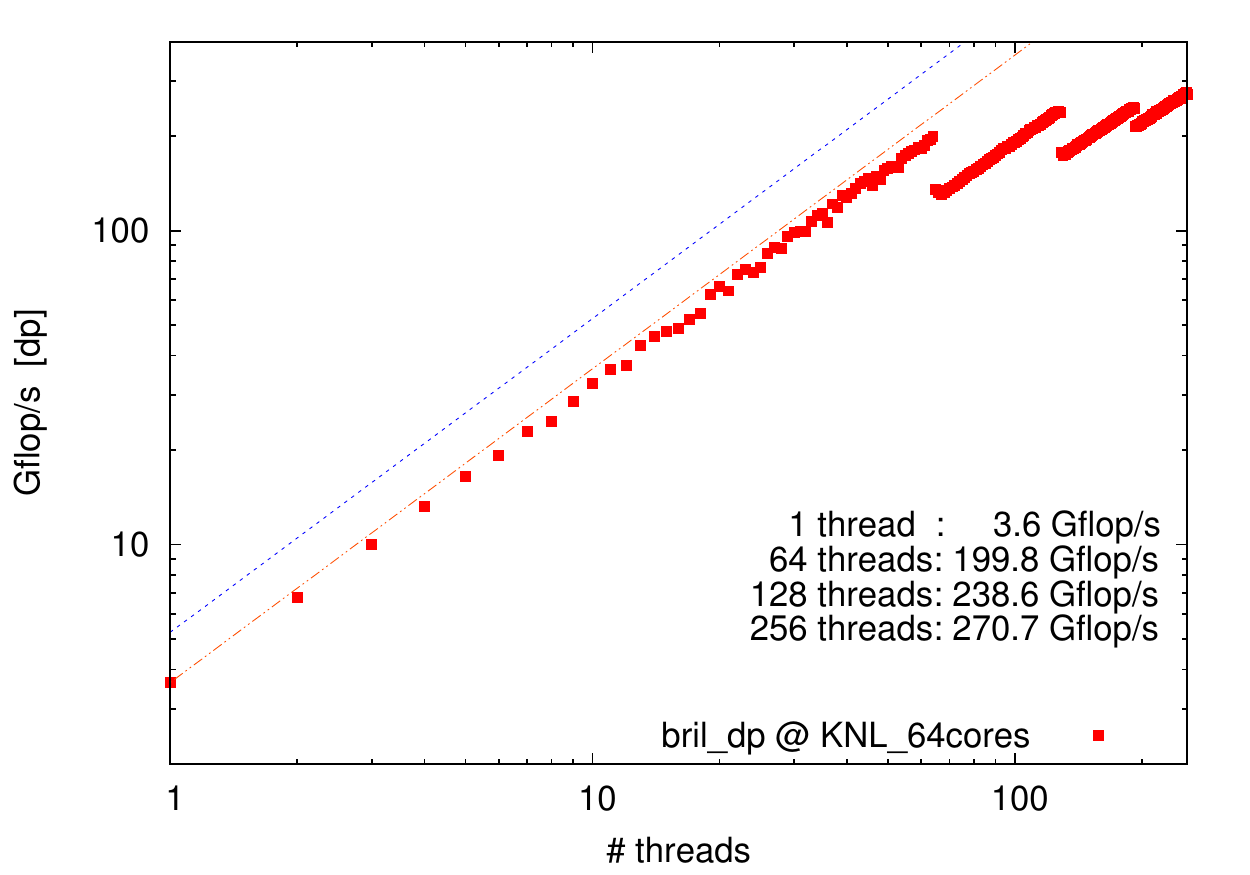}%
\label{fig:bril_scaling}%
\caption{Scaling in the number of threads of the matrix-times-vector performance of the Brillouin operator
on a $32^3\times64$ lattice in {\tt sp} (left) and {\tt dp} (right), using a KNL chip with 64 cores.}%
\end{figure}

The scaling in the number of threads is shown in Fig.\,\ref{fig:bril_scaling}.
We see an almost perfectly linear behavior up to 64 threads; this means that there is essentially no scheduling overhead.
Beyond 64 threads we see load balancing issues among the various threads, except for multiples of 64, where each core is kept busy with exactly the same number of threads.

\begin{table}[tb]
\small
\centering
\begin{tabular}{|c|ccccc|}
\hline
 & $N_\mr{thr}=1$ & $N_\mr{thr}=64$ & $N_\mr{thr}=128$ & $N_\mr{thr}=256$ & $N_\mr{thr}=512$ \\
\hline
{\tt sp} & 5.2 & 301.4 & 326.4 & 356.7 & 357.1 \\
{\tt dp} & 3.6 & 199.8 & 238.6 & 270.7 & 271.8 \\
\hline
\end{tabular}\\[2mm]
\begin{tabular}{|c|cccccc|}
\hline
 & $N_\mr{thr}=1$ & $N_\mr{thr}=2$ & $N_\mr{thr}=4$ & $N_\mr{thr}=6$ & $N_\mr{thr}=12$ & $N_\mr{thr}=24$ \\
\hline
{\tt sp} & 23.6 & 46.0 & 94.8 & 140.3 & 160.1 & 159.2 \\
{\tt dp} & 14.6 & 28.9 & 56.3 &  79.7 &  91.1 &  92.3 \\
\hline
\end{tabular}
\caption{Performance in Gflop/s of the Brillouin matrix-times-vector operation on a $32^3\times64$ lattice,
with $\Nc=3$ and $\Nv=4\Nc$, versus the number of threads. The panels refer to a KNL and a Core i7 (Broadwell), respectively.}%
\label{tab:bril_timing}%
\end{table}

\begin{table}[tb]
\small
\centering
\begin{tabular}{|c|ccccc|}
\hline
{} & $L=12$ & $L=16$ & $L=20$ & $L=24$ & $L=32$ \\
\hline
128 threads & 335 & 328 & 339 & 340 & 321 \\
256 threads & 350 & 350 & 360 & 359 & 357 \\
\hline
\end{tabular}\\[2mm]
\begin{tabular}{|c|cccc|}
\hline
{} & $\Nv=2\Nc$ & $\Nv=4\Nc$ & $\Nv=8\Nc$ & $\Nv=32\Nc$ \\
\hline
128 threads & 194 & 341 & 526 & 533 \\
256 threads & 209 & 360 & 558 & 588 \\
\hline
\end{tabular}\\[2mm]
\begin{tabular}{|c|ccccc|}
\hline
{} & $\Nc=2$ & $\Nc=3$ & $\Nc=4$ & $\Nc=5$ & $\Nc=6$ \\
\hline
128 threads & 344 & 341 & 552 & 444 & 634 \\
256 threads & 356 & 360 & 641 & 488 & 779 \\
\hline
\end{tabular}
\caption{Performance in Gflop/s of the Brillouin matrix-times-vector operation on the KNL architecture in sp arithmetics.
In the first panel the volume dependence is displayed (with $\Nc=3$, $\Nv=4\Nc$, and $T=2L$),
in the second panel the scaling in the number of right-hand-sizes is shown (with $\Nc=3$ and $24^3\times48$ volume),
and in the third panel the dependence on $\Nc$ is considered (with $\Nv=4\Nc$ and $24^3\times48$ volume).}%
\label{tab:bril_extra}%
\end{table}

Some more details are presented in Tab.\,\ref{tab:bril_timing}.
We see some mild improvement when going from 2 to 4 threads per core; beyond 256 threads the performance plateaus.
When comparing sp to dp figures, one should keep in mind that the objects \texttt{W} and \texttt{F} are always in single precision (occupying 5760\,MB and 864\,MB, respectively),
only the vectors \texttt{old} and \texttt{new} change from sp to dp.

Perhaps the most surprising observation is that the same routine (compiled with \texttt{-xcore-avx2} instead of \texttt{-xmic-avx512}) performs well on a standard Core i7 architecture (Broadwell with 6 cores).
Here the plateauing effect sets in after each physical core is occupied with 2 threads.

Some more experiments on the KNL architecture in sp arithmetics are reported in Tab.\,\ref{tab:bril_extra}.
The first panel demonstrates that the $\sim\!360$\,Gflop/s are more or less independent of the volume, i.e.\ we do not see any peculiar cache size effects.
The second panel shows that using more than 12 right-hand-sides (for $\Nc=3$) improves the performance, but beyond 24 right-hand-sides benefits become marginal.
Finally, increasing the number of colors beyond 3 is found to be particularly beneficial.
My personal guess is that with $\Nc=4$ (or a multiple thereof) the colormatrix-times-spinor multiplication in the SIMD loop in Fig.\,\ref{fig:bril_detail} becomes particularly efficient due to a better filling of the SIMD pipeline.

%%%%%%%%%%%%%%%%%%%%%%%%%%%%%%%%%%%%%%%%%%%%%%%%%%%%%%%%%%%%%%%%%%%%%%%%%%%%%

\section{Wilson operator timings}\label{wils-timings}

Timings are done on a node containing a single KNL chip with 64 cores.
All results for the Wilson operator are converted into Gflop/s, based on a flop count of $128\Nc^2+72\Nc$ per site (i.e.\ 1368 for QCD, using the ``shrink-expand-trick'', see e.g.\ \cite{Durr:2017wfi} for details).
The default setup is again a $32^3\times64$ lattice, with $\Nc=3$ and $\Nv=4\Nc$.
This time a sp-field \texttt{V}, a sp-field \texttt{F}, and the dp-fields \texttt{old,new} fit well into the high-bandwidth MCDRAM.
For any given number of threads, all arrays are allocated afresh, and a first touch policy is used as in the Brillouin case.

\begin{figure}[tb]
\includegraphics[width=0.5\textwidth]{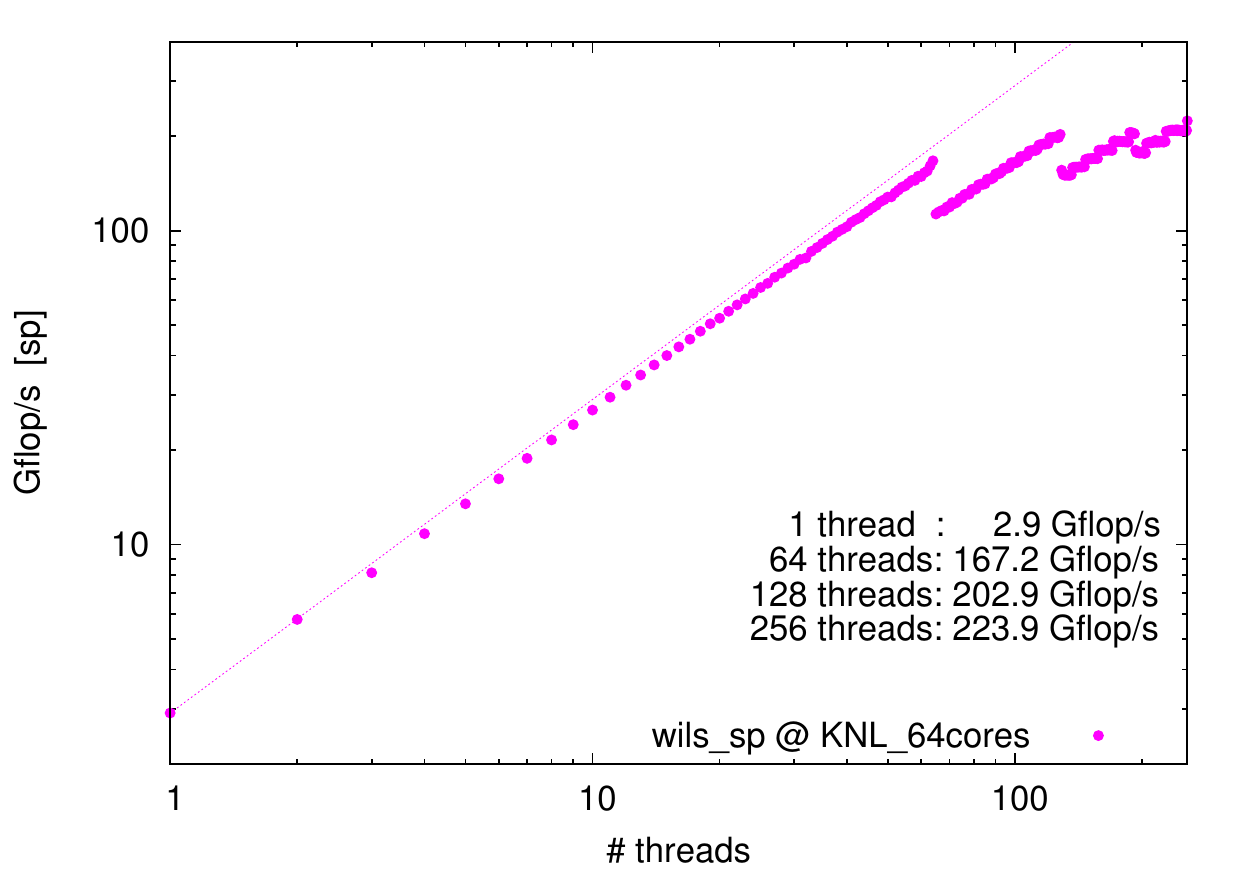}%
\includegraphics[width=0.5\textwidth]{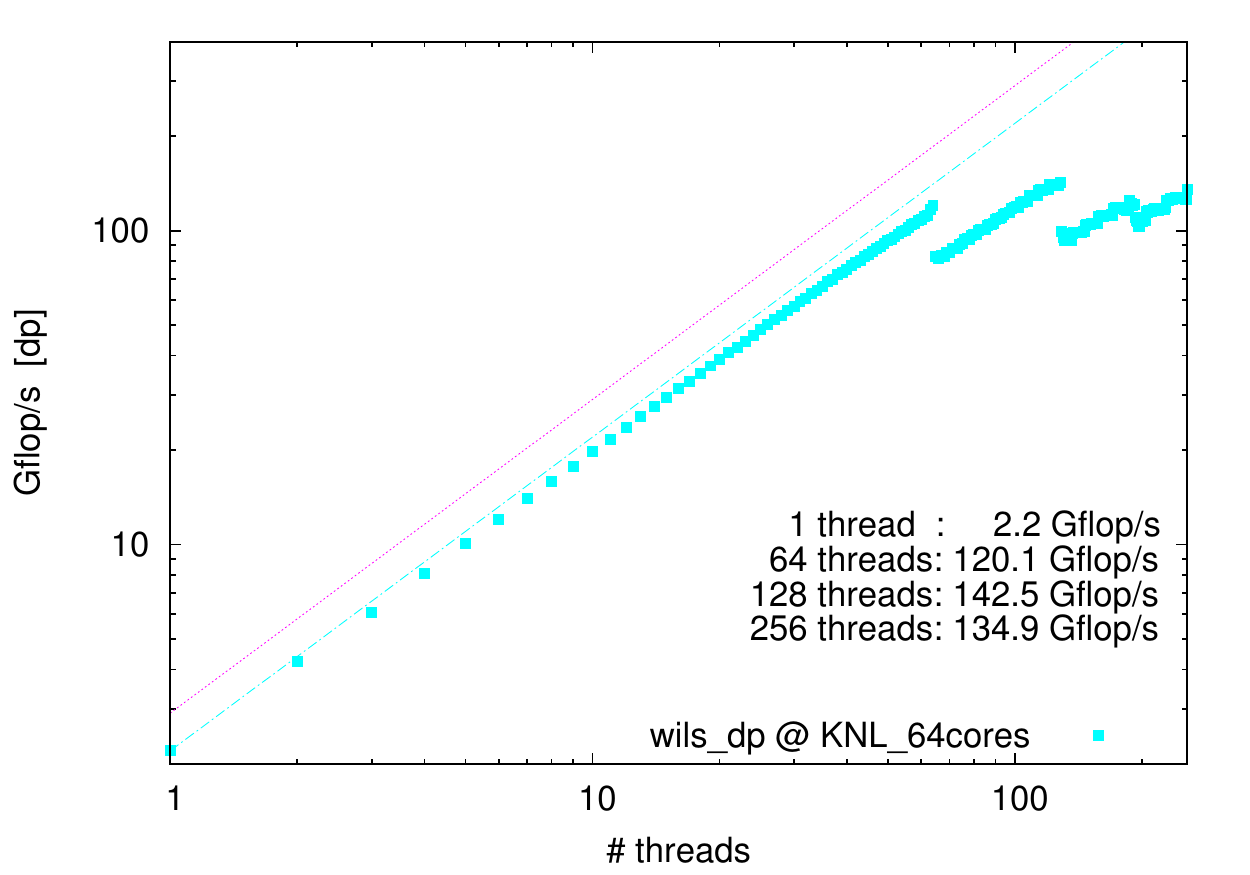}%
\caption{Scaling in the number of threads of the matrix-times-vector performance of the Wilson operator
on a $32^3\times64$ lattice in {\tt sp} (left) and {\tt dp} (right), using a KNL chip with 64 cores.}%
\label{fig:wils_scaling}%
\end{figure}

The scaling in the number of threads is shown in Fig.\,\ref{fig:wils_scaling}.
Again, we see a linear behavior up to 64 threads; and beyond that local maxima are seen for multiples of 64 threads where each core is kept busy with exactly the same number of threads.

\begin{table}[tb]
\small
\centering
\begin{tabular}{|c|ccccc|}
\hline
 & $N_\mr{thr}=1$ & $N_\mr{thr}=64$ & $N_\mr{thr}=128$ & $N_\mr{thr}=256$ & $N_\mr{thr}=512$ \\
\hline
{\tt sp} & 2.9 & 167.2 & 202.9 & 223.9 & 225.2 \\
{\tt dp} & 2.2 & 120.1 & 142.5 & 134.9 & 134.2 \\
\hline
\end{tabular}\\[2mm]
\begin{tabular}{|c|cccccc|}
\hline
 & $N_\mr{thr}=1$ & $N_\mr{thr}=2$ & $N_\mr{thr}=4$ & $N_\mr{thr}=6$ & $N_\mr{thr}=12$ & $N_\mr{thr}=24$ \\
\hline
{\tt sp} & 16.1 & 30.8 & 56.0 & 65.5 & 73.2 & 73.2 \\
{\tt dp} &  9.8 & 18.0 & 32.5 & 39.9 & 38.7 & 38.6 \\
\hline
\end{tabular}
\caption{Performance in Gflop/s of the Wilson matrix-times-vector operation on a $32^3\times64$ lattice,
with $\Nc=3$ and $\Nv=4\Nc$, versus the number of threads. The panels refer to a KNL and a Core i7 (Broadwell), respectively.}%
\label{tab:wils_timing}%
\end{table}

Some more details are presented in Tab.\,\ref{tab:wils_timing}.
On the KNL some some mild improvement is seen when going from 2 to 4 threads per core; beyond 256 threads the performance plateaus.
When comparing sp to dp figures, one should keep in mind that the objects \texttt{V} and \texttt{F} are always in single precision (occupying 576\,MB and 864\,MB, respectively),
only the vectors \texttt{old} and \texttt{new} change from sp to dp.
From the second panel we learn that also the Wilson routine performs (without any change) quite well an the standard Core i7 architecture (Broadwell with 6 cores).
The main difference to the KNL case is that optimum performance is reached with 1 to 2 threads per core rather than 4.

%%%%%%%%%%%%%%%%%%%%%%%%%%%%%%%%%%%%%%%%%%%%%%%%%%%%%%%%%%%%%%%%%%%%%%%%%%%%%

\section{Summary}\label{summary}

The goal of this contribution was to explore whether acceptable performance figures for the Brillouin and Wilson matrix-times-vector applications%
\footnote{After the conference a staggered routine, built with the same guidelines, was found to yield over 280 Gflop/s on the KNL.}
on one KNL chip can be obtained, if we refrain from using advanced optimization techniques (for an overview see the recent plenary talks \cite{Boyle:2017vhi,Rago:2017pyb}).

\begin{table}[tb]
\small
\centering
\begin{tabular}{|l|cc|}
\hline
          & KNL (64 cores)                   & Core i7 (Broadwell)             \\
\hline
Brillouin & 357/272 Gflop/s mean 6.8/10.4\% & 160/92 Gflop/s mean 23.2/26.7\% \\
Wilson    & 225/135 Gflop/s mean 4.3/ 5.2\% & 73 /40 Gflop/s mean 10.6/11.6\% \\
\hline
\end{tabular}
\caption{Conversion of the performance measurements into sustained percentage figures, based on a peak performance of 5.2/2.6 [sp/dp] Tflop/s on the KNL and 690/345 Gflop/s on the Core i7 (Broadwell) architecture.}%
\label{tab:summary}%
\end{table}

Pertinent results are summarized in Tabs.\,\ref{tab:bril_timing} and \ref{tab:wils_timing} for the two operators, respectively.
Not just beauty, also judgement of such figures is in the eye of the beholder.
To me it appears that these are acceptable figures -- especially in view of the simplicity of the shared memory parallelization and SIMD encouragement strategies used (both with OMP pragmas only).

Perhaps the most surprising finding is that these routines (unchanged, just recompiled) perform quite well on the standard Core i7 architecture, too.
The loss in performance, compared to the KNL architecture, is a factor 2.2 for the Brillouin operator, and a factor 3.1 for the Wilson operator.

%%% KNL64: 1.269*64*32*[2,1]=[5.2,2.6] [Tflop/s]
%%% Broad: 3.600*06*16*[2,1]=[690,345] [Gflop/s]

It is instructive to convert these figures into sustained performance ratios.
The KNL chip operates at 1.269\,GHz; with 64 cores and 64/32 flop/s per cycle it has a peak performance of 5.2/2.6\,Tflop/s in sp/dp arithmetics.
The Broadwell chip operates at 3.6\,GHz; with 6 cores and 32/16 flop/s per cycle it has a peak performance of 690/345\,Gflop/s in sp/dp arithmetics.
With these numbers in hand, the performance figures of the Brillouin and Wilson operators can be converted into sustained performance ratios.
The results, collected in Tab.\,\ref{tab:summary}, indicate that (a) the efficiency of the Brillouin operator is generically higher than the efficiency of the Wilson operator,
and (b) the efficiency on the Broadwell architecture is generically higher than the efficiency of the KNL.

An explanation of (a) is easily found.
For SU(3) gauge group the Brillouin-to-Wilson flop-count ratio is 22.1.
At the same time the Brillouin-to-Wilson memory-traffic ratio is 8.9 \cite{Durr:2017wfi}.
Taken together, this means that the \emph{computational intensity} of the Brillouin operator is higher by a factor 2.5 \cite{Durr:2017wfi}.
We know that modern architectures tend to have plenty of CPU capability, and such prerequisites favor applications with high computational intensity.
As for (b) the overall time (as compared to e.g.\ a scalar product) and the huge performance difference between SU(3) and SU(4) gauge group suggest that incomplete filling of the SIMD pipeline in the \texttt{idx}-loops in Figs.\,\ref{fig:wils_detail} and \ref{fig:bril_detail} likely represents the actual bottleneck (at least on the KNL architecture which operates at 512-bit width).

The main lesson is that in Lattice QCD it is easy to get a reasonable (i.e.\ non-excellent) performance, while maintaining full portability, if the compiler acts on code whose structure is \emph{simple}%
\footnote{I thank Eric Gregory and Christian Hoelbling for discussion; I acknowledge partial funding by DFG through SFB TR-55.}.

%%%%%%%%%%%%%%%%%%%%%%%%%%%%%%%%%%%%%%%%%%%%%%%%%%%%%%%%%%%%%%%%%%%%%%%%%%%%%

\bibliography{Lattice2017}

%%%%%%%%%%%%%%%%%%%%%%%%%%%%%%%%%%%%%%%%%%%%%%%%%%%%%%%%%%%%%%%%%%%%%%%%%%%%%

\end{document}